\documentclass[epj,twocolumn]{webofc} 
\usepackage[varg]{txfonts}   
 \woctitle{TRANSVERSITY 2014} %

\newcommand{\bgc}{\begin{center}}

\newcommand{\be}{\begin{equation}}
\newcommand{\ee}{\end{equation}}
\newcommand{\ben}{\begin{eqnarray}}
\newcommand{\een}{\end{eqnarray}}
\def\bea#1\eea{\begin{align}#1\end{align}}

\newcommand{\bef}{\begin{figure}[htb]\centering}
\newcommand{\eef}{\end{figure}}

\newcommand{\bfl}{\begin{flushleft}}
\newcommand{\efl}{\end{flushleft}}

\newcommand{\Php}{P_{h\perp}}

\newcommand{\enc}{\end{center}}

\newcommand{\bgi}{\begin{itemize}}
\newcommand{\eni}{\end{itemize}}

\newcommand{\bfk}{\mbox{\boldmath $k$}}

\newcommand{\bfp}{\mbox{\boldmath $p$}}

\newcommand{\bfP}{\mbox{\boldmath $P$}}
\newcommand{\bfS}{\mbox{\boldmath $S$}}
\newcommand{\bfs}{\mbox{\boldmath $s$}}

\newcommand{\pup}{p^\uparrow}

\newcommand{\qup}{q^\uparrow}

\def\lsim{\mathrel{\rlap{\lower4pt\hbox{\hskip1pt$\sim$}}\raise1pt\hbox{$<$}}}
\def\gsim{\mathrel{\rlap{\lower4pt\hbox{\hskip1pt$\sim$}}\raise1pt\hbox{$>$}}}
\def\nostrocostruttino#1\over#2{\mathrel{\mathop{\kern 0pt \rlap
{\hbox{$#1$}}} \hbox{\kern-.135em $#2$}}}

\begin{document}
\title{$A_N$ in inclusive lepton-proton collisions: TMD and twist-3 approaches}
%
%

\author{Alexei Prokudin\inst{1}\fnsep\thanks{\email{prokudin@jlab.org}} }

\institute{Jefferson Lab, 12000 Jefferson Avenue, Newport News, VA 23606, USA}

\abstract{%
 We consider the inclusive production of hadrons in lepton-nucleon scattering.
For a transversely polarized nucleon this reaction shows a left-right azimuthal asymmetry, which we compute in both TMD and in twist-3 collinear factorization formalisms. 
All non-perturbative parton correlators of the calculation are fixed through information from other hard processes.
Our results for the left-right asymmetry agree in sign and size with the HERMES Collaboration and Jefferson Lab recent data for charged pion production.
We discuss similarities and differences of the two formalisms.
}
\maketitle
\section{Introduction}
\label{intro}
These proceedings are based on two papers published in collaboration with  
Mauro Anselmino, Mariaelena Boglione,
Umberto D'Alesio, 
Stefano Melis,  
Francesco Murgia in Ref.~\cite{Anselmino:2014eza}
 and with Leonard Gamberg, 
 Zhong-Bo Kang, Andreas Metz, Daniel Pitonyak in Ref.~\cite{Gamberg:2014eia}. All results in these proceedings follow Refs.~\cite{Anselmino:2014eza,Gamberg:2014eia}.

Let us consider inclusive production of hadrons in lepton-nucleon scattering, $\ell \, N \to h \, X$.
If the transverse momentum $P_{h\perp}$ of the final state hadron is sufficiently large, this presents a very interesting testing ground of two different but related factorization schemes. One is the so-called Transverse Momentum Dependent factorization and the other twist-3 collinear factorization.
 
Our focus here is on the left-right azimuthal asymmetry that can be defined if the nucleon is transversely polarized.
This asymmetry is similar to the transverse single-spin asymmetry $A_N$ which has already been studied extensively in hadronic collisions like $p^{\uparrow} p \to h \, X$ --- see Refs.~\cite{Adams:1991cs,Adler:2005in,Adare:2013ekj,Adare:2014qzo,Adams:2003fx,Abelev:2008qb,Adamczyk:2012xd,Arsene:2008mi,Bland:2013pkt,Efremov:1981sh,Efremov:1984ip,Qiu:1991pp,Qiu:1991wg,Qiu:1998ia,Kouvaris:2006zy,Kang:2011hk,Anselmino:1994tv,Anselmino:1998yz,Anselmino:2005sh,Anselmino:2012rq,Anselmino:2013rya,Kanazawa:2000hz,Eguchi:2006mc,Koike:2009ge,Kanazawa:2010au,Kanazawa:2011bg,Beppu:2013uda,Kang:2010zzb, Metz:2012ui, Metz:2012ct, Kanazawa:2014dca}.
Recently, the HERMES Collaboration~\cite{Airapetian:2013bim} and the Jefferson Lab Hall A Collaboration~\cite{Allada:2013nsw} reported the first ever measurements of $A_N$ in lepton-nucleon scattering.
In general, one may expect that $A_N$ in this reaction could give new insight into the underlying mechanism of $A_N$ in hadronic collisions, which is the subject of longstanding discussions.

In twist-3 factorization one computes \cite{Gamberg:2014eia} $A_N$ using  two main components:
First, a twist-3 effect originates from the transversely polarized nucleon.
In that case the key non-perturbative entity is the so-called Qiu-Sterman (QS) function~\cite{Qiu:1991pp,Qiu:1991wg} --- a specific quark-gluon-quark correlator that has an intimate connection with the transverse momentum dependent (TMD) Sivers function~\cite{Sivers:1989cc,Sivers:1990fh}.
Second, a twist-3 effect also arises from parton fragmentation.
This contribution can be expressed by means of two independent fragmentation correlators~\cite{Yuan:2009dw,Metz:2012ct,Kanazawa:2013uia}, one of which is related to the Collins fragmentation function~\cite{Collins:1992kk}.
A first attempt to get a complete result for $A_N$ in $\ell \, p^{\uparrow} \to h \, X$ in the collinear twist-3 approach can be found in a conference proceeding~\cite{Koike:2002ti}.

On the other hand one could assume TMD factorization and then compute the asymmetry using TMD functions, as done in \cite{Anselmino:1999gd,Anselmino:2009pn,Anselmino:2014eza}.  Assuming the validity of the TMD factorization scheme for the process
$p \,\ell \to h \, X$ in which the only large scale detected is the transverse
momentum $P_T$ of the final hadron in the proton-lepton {\it c.m.} frame, the
main contribution to $A_N$ comes from the Sivers and Collins
effects ~\cite{Anselmino:2009pn,D'Alesio:2004up,Anselmino:2005sh,D'Alesio:2007jt}.

We have studied~\cite{Anselmino:2014eza,Gamberg:2014eia} the process in these two frameworks and report here results and comparison.

\section{Kinematics and analytical results in twist-3 approach \label{sectionII}}
\noindent
Here we discuss some details of the kinematics and present the tree level results for the unpolarized and the spin-dependent cross section entering the definition of $A_N$. 
For the process under consideration 
\bea
\ell (l)  + N(P, S_P) \to h(P_h) + X \,,
\eea
$l$, $P$, and $P_h$ denote the momentum of the lepton, nucleon, and produced hadron, respectively, and $S_P$ is the spin vector of the nucleon. 
We use the momenta of the particles to fix a coordinate system according to $\hat{e}_z = \hat{P} = - \hat{l}$, $\hat{e}_x = \hat{P}_{h\perp}$, and $\hat{e}_y = \hat{e}_z \times \hat{e}_x$.
The Mandelstam variables for the scattering process are defined by
\begin{equation} \label{e:mandel_1}
S = (l + P)^2 \,, \qquad
T = (P - P_h)^2 \,, \qquad 
U = (l - P_h)^2 \,,
\end{equation}
while at the corresponding  partonic level one has
\begin{equation} \label{e:mandel_2}
\hat{s} = (l + k)^2 = x S\,, \;
\hat{t} = (k - p)^2 = \frac{x T}{z} \,, \;
\hat{u} = (l - p)^2 = \frac{U}{z} \,,
\end{equation}
with $k$ characterizing the momentum of the active quark in the nucleon, and $p$ the momentum of the fragmenting quark. 
Neglecting parton transverse momenta one has $k = x P$ and $p = P_h / z$.

For the unpolarized lepton-nucleon collisions, the differential cross section at leading order (LO) is given by~\cite{Kang:2011jw}
\begin{align}
P_h^0 \, \frac{d\sigma_{UU}} {d^3\vec{P}_h} = \frac{2\alpha_{\rm em}^2} {S} \,
\sum_q e_q^2 \int_{z_{\rm min}}^1 \frac{dz} {z^2} \, \frac{1} {S+T/z} \, \nonumber \\ 
\times \frac{1} {x} \, f_1^q(x) \, D_1^{h/q}(z) 
\bigg[ \frac{\hat{s}^2 + \hat{u}^2}{\hat{t}^2} \bigg] \,,
\label{e:lNhXUU}
\end{align}
where $f_1^q$ is the unpolarized quark distribution, and $D_1^{h/q}$ is the unpolarized fragmentation function.
Here $z_{\rm min}= -(T+U)/S$, and $x$ can be determined from the on-shell condition $\hat s+ \hat t + \hat u=0$ in our LO formula as
\bea
x=-(U/z)/(S+T/z) \,.
\label{e:xdef}
\eea
We now turn to the spin-dependent cross section for the process $\ell \, N^{\uparrow}  \to h \, X$, that is,  an unpolarized lepton scattering off 
a transversely polarized nucleon. 
We work in the collinear factorization framework, in which this cross section is a twist-3 observable. 
The twist-3 effect can either come from the side of the parton distribution in the transversely polarized nucleon~\cite{Qiu:1991pp}, or from the side of the parton fragmentation into the final-state hadron~\cite{Yuan:2009dw,Kang:2010zzb,Metz:2012ct}. 
Calculations for such a twist-3 observable in collinear factorization have become standard, and details can be found in the literature --- see, e.g., Refs.~\cite{Qiu:1991pp,Qiu:1991wg,Qiu:1998ia,Kouvaris:2006zy,Kanazawa:2000hz,Eguchi:2006mc,Koike:2009ge,Kang:2010zzb,Metz:2012ct,Yuan:2009dw,Kanazawa:2013uia,Kang:2008qh,Kang:2008ih,Zhou:2009jm,Beppu:2010qn,Koike:2011mb,Liang:2012rb,Metz:2012fq,Hatta:2013wsa}. In particular, we refer to~\cite{Metz:2012ct} where the fragmentation contribution to $A_N$ for $p^{\uparrow} p \to h \, X$ has been computed.
Here we only write down the final expression 
\begin{align}
& P_h^0 \,\frac{ d\sigma_{UT}} {d^3\vec{P}_h}  =  - \frac{8\alpha_{\rm em}^2} {S} \,
\varepsilon_{\perp\mu\nu} \, S_{P\perp}^{\mu} \, P_{h\perp}^{\nu} \,  
\nonumber \\ 
 & \times \sum_q e_q^2 \int_{z_{\rm min}}^1 \frac{dz} {z^3}\,\frac{1} {S+T/z}\,\frac{1} {x} 
\nonumber \\
&  \times \Bigg\{\!\!-\!\frac{\pi M} {\hat{u}}\,D_1^{h/q}(z) \bigg(F_{FT}^q(x,x)-x\frac{dF_{FT}^q(x,x)} {dx}\bigg) 
  \nonumber \\
 & \times\!\bigg[\frac{\hat{s}(\hat{s}^2+\hat{u}^2)} {2\hat{t}^{\hspace{0.025cm}3}}\bigg] 
\nonumber \\
& +\,\frac{M_h} {-x\hat{u}-\hat{t}}\,\,h_{1}^{q}(x)\,\Bigg\{\!\!\bigg(\hat{H}^{h/q}(z)-z\frac{d\hat{H}^{h/q}(z)} {dz}\bigg)\!\bigg[\frac{(1-x)\hat{s}\hat{u}} {\hat{t}^{\hspace{0.025cm}2}}\bigg]  
\nonumber \\
& +\, \frac{1} {z} \, H^{h/q}(z) \bigg[ \frac{\hat{s} (\hat{s}^2 +(x-1)\hat{u}^2)} {\hat{t}^{\hspace{0.025cm}3}}\bigg] 
\nonumber \\
& + 2 z^2 \! \int_z^\infty \! \frac{dz_1} {z_1^2} \, \frac{1} {\frac{1} {z} -\frac{1} {z_{1}}} \, \hat{H}_{FU}^{h/q,\Im}(z,z_{1}) 
\bigg[ \frac{x\hat{s}^2\hat{u}} {\xi_{\hspace{0.025cm}z} \, \hat{t}^{\hspace{0.025cm}3}} \bigg]\!\Bigg\}\!\Bigg\} \,, 
\label{e:lNhXUT}
\end{align}

where we use the convention $\varepsilon^{12}_{\perp} \equiv \varepsilon^{-+12} = 1$, and $\xi_{\,z} = z/z_g$ with $1/z_{g}=(1/z-1/z_1)$.  At the operator level and in a parton model analysis, the QS function $F_{FT}^q$~\cite{Qiu:1991pp,Qiu:1991wg} can be related to the first $k_\perp$ moment of the Sivers function $f_{1T}^{\perp q}$~\cite{Boer:2003cm,Kang:2011hk},
\be
\pi \, F_{FT}^q(x,x) = \int d^2 \vec{k}_{\perp} \, \frac{\vec{k}_{\perp}^{\,2}}{2 M^2} \, f_{1T}^{\perp q}(x,\vec{k}_{\perp}^{\,2}) \Big|_{\rm SIDIS} \,,
\label{e:ETQS}
\ee
where the subscript ``SIDIS'' indicates the Sivers function probed in semi-inclusive deep-inelastic scattering. 
The function $\hat{H}^{h/q}$ has the following relation to the Collins function $H_1^{\perp h/q}$~\cite{Yuan:2009dw,Kang:2010zzb,Metz:2012ct},
\be
\hat{H}^{h/q}(z) = z^2 \int d^2 \vec{p}_{\perp} \, \frac{\vec{p}_{\perp}^{\,2}}{2 M_h^2} \, H_1^{\perp h/q}(z,z^2\vec{p}_{\perp}^{\,2}) \,.
\label{e:HMOM}
\ee
Our definitions for both $f_{1T}^{\perp q}$ and $H_1^{\perp h/q}$ follow the so-called Trento convention \cite{Bacchetta:2004jz}. 
On the fragmentation side $\sigma_{UT}$ contains two additional twist-3 terms.
Those depend on the two-parton correlator $H^{h/q}$ and the (imaginary part of the) 3-parton correlator $\hat{H}_{FU}^{h/q}$. The underlying dynamics for these functions may be similar to the one for the Collins effect, and it turns out in fact that $\hat{H}^{h/q}, H^{h/q}$, and $\hat{H}_{FU}^{h/q,\Im}$ are not independent of each other but satisfy the relation~\cite{Metz:2012ct}
\be
H^{h/q}(z) = -2 z  \hat H^{h/q}(z)+ 2 z^3 \! \int_z^\infty \! \frac{dz_1} {z_1^2} \, \frac{1} {\frac{1} {z} -\frac{1} {z_{1}}} \, \hat{H}_{FU}^{h/q,\Im}(z,z_{1}) \,.
\label{e:relation}
\ee
 
\section{Analytical results in TMD approach\label{sectionIII}}
In Ref.~\cite{Anselmino:2009pn} (to which we refer for all details) we
considered the process $\pup \ell \to h \, X$ in the proton-lepton {\it c.m.}
frame (with the polarised proton moving along the positive $Z_{cm}$ axis)
and the transverse single spin asymmetry:
\be
A_N = \frac{d\sigma^\uparrow(\bfP_T) - d\sigma^\downarrow(\bfP_T)}
           {d\sigma^\uparrow(\bfP_T) + d\sigma^\downarrow(\bfP_T)}
    = \frac{d\sigma^\uparrow(\bfP_T) - d\sigma^\uparrow(-\bfP_T)}
           {2 \, d\sigma^{\rm unp}(\bfP_T)} \,, \label{an}
\ee
where
\be
d\sigma^{\uparrow, \downarrow} \equiv \frac{E_h \, d\sigma^{p^{\uparrow,
\downarrow} \, \ell \to h\, X}}{d^{3} \bfP_h}
\ee
is the cross section for the inclusive process $p^{\uparrow, \downarrow}
\, \ell \to h \, X$ with a transversely polarised proton with spin ``up"
($\uparrow$) or ``down" ($\downarrow$) with respect to the scattering
plane~\cite{Anselmino:2009pn}.
$A_N$ can be measured either by looking at the production of hadrons
at a fixed transverse momentum $\bfP_T$, changing the incoming proton
polarization from $\uparrow$ to $\downarrow$, or keeping a fixed
proton polarization and looking at the hadron production to the left
and the right of the $Z_{cm}$ axis (see Fig.~1 of Ref.~\cite{Anselmino:2009pn}). $A_N$ was defined (and computed)
for a proton in a pure spin state with a pseudo-vector polarization
$\bfS_T$ normal ($N$) to the production plane and $|\bfS_T| = S_T = 1$.
For a generic transverse polarization along an azimuthal direction $\phi_S$
in the chosen reference frame, in which the $\uparrow$ direction is given by
$\phi_S = \pi/2$, and a polarization $S_T \not= 1$, one has:
\be
A(\phi_S, S_T) = \bfS_T \cdot (\hat{\bfp} \times \hat{\bfP}_T) \, A_N =
S_T \sin\phi_S \, A_N \>, \label{phis}
\ee
where $\bfp$ is the proton momentum. Notice that if one follows the usual
definition adopted in SIDIS experiments, one simply has:
\be
A_{TU}^{\sin\phi_S} \equiv \frac{2}{S_T} \,
\frac{\int \, d\phi_S \> [d\sigma(\phi_S) - d\sigma(\phi_S + \pi)]\> \sin\phi_S}
     {\int \, d\phi_S \> [d\sigma(\phi_S) + d\sigma(\phi_S + \pi)]}
= A_N \>.
\label{ATU}
\ee
The asymmetry can be written as:
\be
A_N =
\frac
{{\displaystyle \sum_{q,\{\lambda\}} \int }\> [\Sigma(\uparrow) - \Sigma(\downarrow)]^{q \ell \to q \ell}}
{{\displaystyle \sum_{q,\{\lambda\}} \int }\> [\Sigma(\uparrow) + \Sigma(\downarrow)]^{q \ell \to q \ell}} \>,
\label{anh}
\ee
where $\int$ stands for 
\be
 \int \frac{dx \, dz}
{16\,\pi^2 x\,z^2 s}
d^2 \bfk_{\perp} \, d^3 \bfp_{\perp}\,
\delta(\bfp_{\perp} \cdot \hat{\bfp}'_q) \, J(p_\perp)
\> \delta(\hat s + \hat t + \hat u)
\ee
and
\bea
\sum_{\{\lambda\}}\,[\Sigma(\uparrow) - \Sigma(\downarrow)]^{q
\ell \to q \ell} = \frac{1}{2} \, \Delta^N\! f_{q/\pup}
(x,k_{\perp}) \cos\phi \nonumber \\
\times  \, \left[\,|{\hat M}_1^0|^2 + |{\hat
M}_2^0|^2 \right] \,
D_{h/q} (z, p_{\perp})  \nonumber \\
+ h_{1q}(x,k_{\perp}) \, \hat M_1^0 \hat M_2^0 \, \Delta^N\!
D_{h/\qup} (z, p_{\perp}) \, \cos(\phi' + \phi_q^h) \label{ds1}
\eea
and
\bea
\sum_{\{\lambda\}}\,[\Sigma(\uparrow) +
\Sigma(\downarrow)]^{q \ell \to q \ell} =
f_{q/p} (x,k_{\perp}) \,
\left[\,|{\hat M}_1^0|^2 + |{\hat M}_2^0|^2 \right] \nonumber \\
\times \,
D_{h/q} (z, p_{\perp}) \>. \label{ss1}
\eea

All functions and all kinematical and dynamical variables appearing in the above equations are exactly defined in Ref.~\cite{Anselmino:2009pn} and its Appendices and in Ref.~\cite{Anselmino:2005sh}.  

TMD functions in notations of Refs.~\cite{Anselmino:2014eza,Gamberg:2014eia} are related and exact relations can be found in so-called Trento convention Ref.~\cite{Bacchetta:2004jz}.

The first term on the r.h.s.~of Eq.~(\ref{ds1}) shows the contribution to $A_N$ of the Sivers function $\Delta^N\!
f_{q/\pup}(x,k_\perp)$~\cite{Sivers:1989cc,Sivers:1990fh,Bacchetta:2004jz},
\bea
\Delta \hat f_{q/p,S}(x, \bfk_{\perp}) = \hat f_{q/p,S}(x,
\bfk_{\perp}) - \hat f_{q/p,-S}(x, \bfk_{\perp}) \nonumber \\ \equiv \Delta^N\!
f_{q/\pup}\,(x, k_{\perp}) \>
\bfS_T \cdot (\hat{\bfp} \times \hat{\bfk}_{\perp }) \label{defsivnoi} \\
= -2 \, \frac{k_\perp}{M} \, f_{1T}^{\perp q}(x, k_{\perp}) \>
\bfS_T \cdot (\hat{\bfp} \times \hat{\bfk}_{\perp }) \>. \nonumber
\eea

The second term on the r.h.s.~of Eq.~(\ref{ds1}) shows the contribution to $A_N$ of the unintegrated transversity distribution $h_{1q}(x,k_{\perp})$ coupled to the Collins function $\Delta^N\! D_{h/\qup} (z, p_{\perp})$~\cite{Collins:1992kk,Bacchetta:2004jz},
\bea
\Delta \hat D_{h/q^\uparrow}\,(z, \bfp_{\perp}) = \hat
D_{h/q^\uparrow}\,(z, \bfp_{\perp}) - \hat D_{h/q^\downarrow}\,(z,
\bfp_{\perp}) \nonumber 
\\ \equiv \Delta^N\! D_{h/\qup}\,(z, p_{\perp}) \>
\bfs_q \cdot (\hat{\bfp}_q^\prime \times \hat{\bfp}_{\perp }) \\
\label{defcolnoi} = \frac{2 \, p_\perp}{z \, m_h} H_{1}^{\perp q}(z,
p_{\perp}) \> \bfs_q \cdot (\hat{\bfp}_q^\prime \times
\hat{\bfp}_{\perp}) \>. \nonumber
\eea

\section{Numerical results and comparison \label{sectionIV}}
\begin{figure}[h!t]
\includegraphics[width=6.truecm,angle=0]{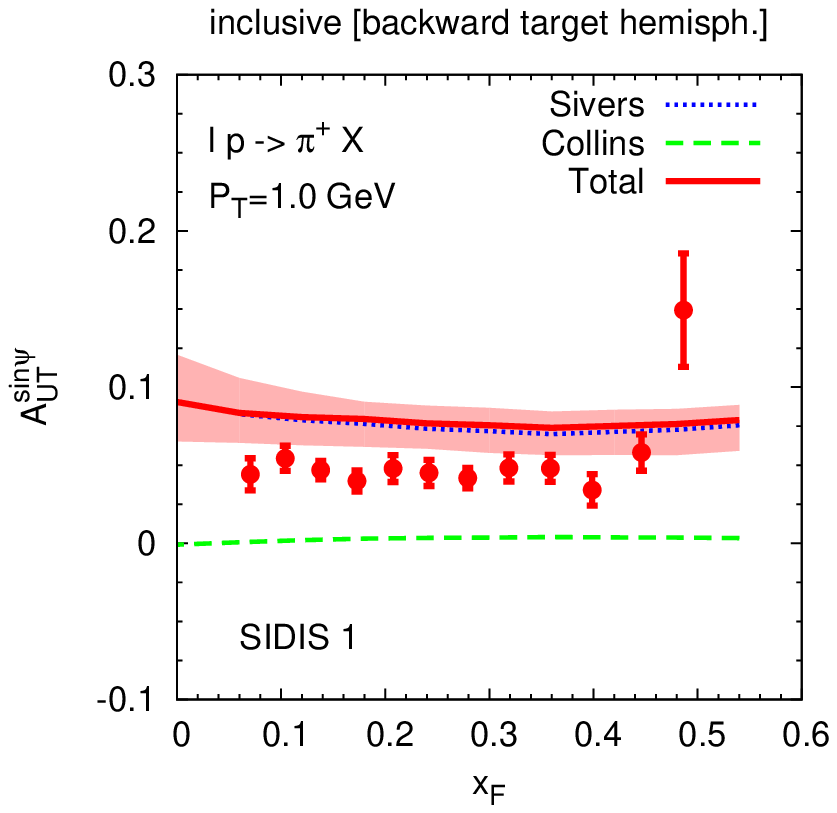}
\includegraphics[width=6.truecm,angle=0]{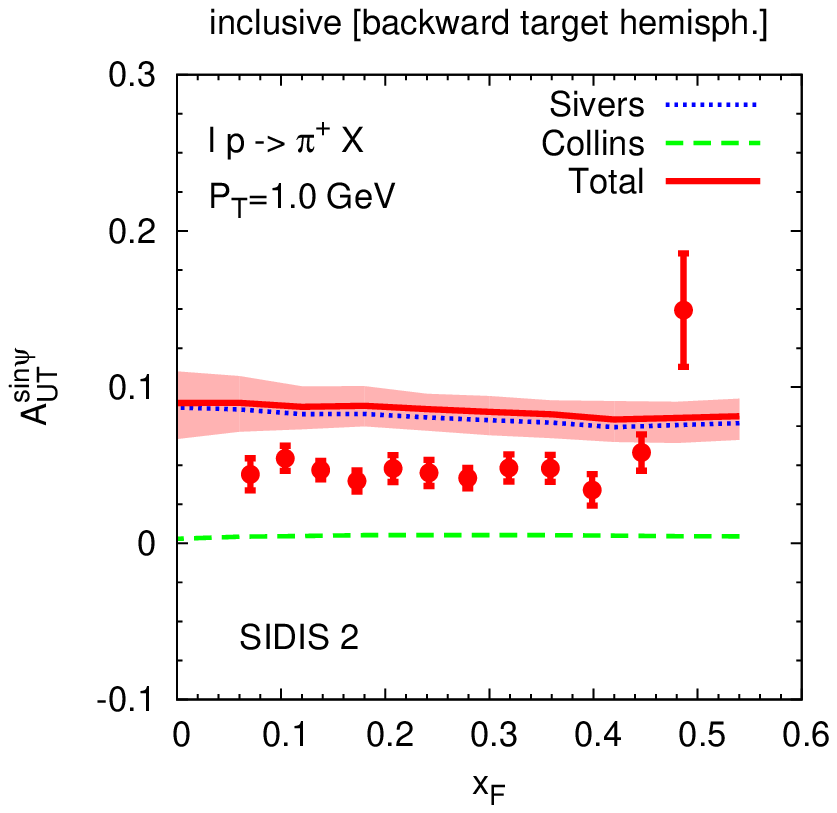}
\caption{The theoretical estimates for $A_{UT}^{\sin\psi}$ vs.~$x_F$ at
$\sqrt{s}\simeq 7$ GeV and $P_T = 1$ GeV for inclusive $\pi^+$ production in
$\ell \, \pup \to \pi \, X$ processes, computed according to
Eqs.~(\ref{anh})--(\ref{ss1}) of the text, are compared
with the HERMES data~\cite{Airapetian:2013bim}. The contributions from the
Sivers (dotted blue lines) and the Collins (dashed green lines) effects are
shown separately and also added together (solid red lines). The computation
is performed adopting the Sivers and Collins functions of
Refs.~\cite{Anselmino:2005ea, Anselmino:2007fs}, referred to as SIDIS~1 in
the text (left panel), and of Refs.~\cite{Anselmino:2008sga, Anselmino:2008jk},
SIDIS~2 in the text (right panel). The overall statistical uncertainty band,
also shown, is the envelope of the two independent statistical uncertainty
bands obtained following the procedure described in Appendix A of
Ref.~\cite{Anselmino:2008sga}.}
\label{fig:an-hermes-pip}
\end{figure}

\noindent
In this case, in order to apply our TMD factorised approach, one has to
consider data at large $P_T$. Among the HERMES data there is one bin that
fulfils this requirement, with $1 \lsim P_T \lsim 2.2$ GeV, and $\langle
P_T \rangle \simeq$ 1--1.1 GeV. In Fig.~\ref{fig:an-hermes-pip}   we show a comparison of our estimates with these data
  for positive   pion production. Sivers functions extracted in Ref.~\cite{Anselmino:2005ea} using  
the Kretzer set for the collinear FFs~\cite{Kretzer:2000yf} is referred to as SIDIS~1 set.
More recent extraction is of Ref.~\cite{Anselmino:2008sga}, where also the sea quark contributions
were included,  we adopted another
set for the FFs, namely that one by de Florian, Sassot and Stratmann (DSS)~\cite{deFlorian:2007aj}. We refer to
these as the SIDIS~2 set.

In this kinematical region the Collins effect is always negligible,
almost compatible with zero. The reason is twofold: first, the partonic
spin transfer in the backward proton hemisphere is dynamically suppressed,
as explained in Ref.~\cite{Anselmino:2009pn}; second, the azimuthal phase
(see the second term on the r.h.s.~of Eq.~(\ref{ds1})) oscillates strongly, washing out the
effect.

The Sivers effect does not suffer from any dynamical suppression, since it
enters with the unpolarised partonic cross section. Moreover, there is no
suppression from the integration over the azimuthal phases, as it happens,
for instance, in $p \, p \to \pi \, X$ case. Indeed in $\ell \, p \to \pi \, X$
only one partonic channel is at work and, for the moderate $Q^2$ values
of HERMES kinematics, the Sivers phase ($\phi$) appearing in the first
term on the r.h.s.~of Eq.~(\ref{ds1}) appears also significantly in the elementary
interaction, thus resulting in a non-zero phase integration.

Moreover, in this kinematical region, even if looking at the backward hemisphere
of the polarised proton, one probes its valence region, where the extracted
Sivers function are well constrained. The reason is basically related to the
moderate {\it c.m.} energy, $\sqrt s \simeq 7$ GeV, of the HERMES experiment.

As one can see, the SSA for positive pion production is a bit overestimated,
Fig.~\ref{fig:an-hermes-pip}.
Notice that in the fully inclusive case under study, at such values of $\sqrt s$
and $Q^2$ other effects could contaminate the SSA. Nonetheless the qualitative description, in size, shape and sign, is quite encouraging.

\begin{figure}[h!t]
\includegraphics[width=8.truecm,angle=0]{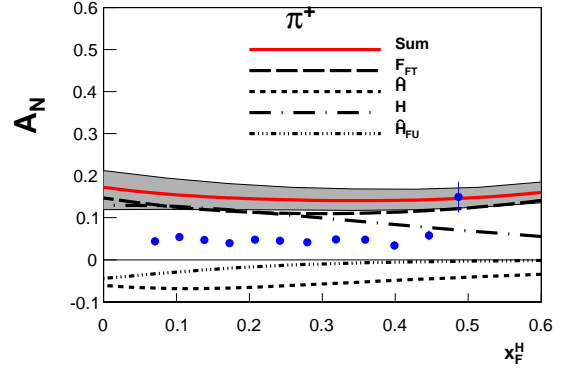}
\caption{$A_N$ as a function of $x_F^H$ for $\pi^+$  production at $\Php = 1 \, \rm{GeV}$ for lepton-proton collisions at $\sqrt{S} = 7.25 \, \rm{GeV}$. 
The data are from Ref.~\cite{Airapetian:2013bim}. 
The solid line corresponds to the sum of all contributions. 
The $F_{FT}^q$ contribution is the dashed line, the $\hat{H}^{h/q}$ contribution is the dotted line, the ${H}^{h/q} $contribution is the dot-dashed line, and the $\hat{H}^{h/q}_{FU}$ contribution is the 3-dotted-dashed line. 
The error band comes from uncertainties in the Sivers, Collins, and transversity functions estimated in Refs.~\cite{Anselmino:2008sga,Anselmino:2013vqa}.  
Note that positive $x_F^{\rm H}$ corresponds to pions in the backward direction with respect to the target proton.}
\label{fig:an_hermes_pipm_xf}
\end{figure}

 Using twist-3 functions we compute asymmetry by Eqs.~(\ref{e:lNhXUU},\ref{e:lNhXUT}). 
 In the following we will plot $A_N (-x_F,\Php) = A_{UT}^{\sin{\Psi}}(x_F^H, \Php)$ as a function of $x_F^H$ and $\Php$.
 In Fig.~\ref{fig:an_hermes_pipm_xf} we plot $A_N$ as a function of $x_F^{\rm H}$ for $\pi^+$   production with $1 < \Php < 2.2 \; \rm{GeV}$ ($\langle \Php\rangle \simeq 1$ GeV) for lepton-proton collisions at HERMES energy $\sqrt{S} = 7.25$ GeV \cite{Airapetian:2013bim}. 
For $\pi^+$ the contribution coming from $F_{FT}^q$ related to the Sivers effect is positive for all $x_F$.  
The contribution from $\hat{H}^{h/q}$ is of opposite sign and smaller in absolute value than that from $F_{FT}^q$.  
The contribution from ${H}^{h/q}$ is positive and that from $\hat{H}^{h/q,\Im}_{FU}$ is negative, and their sum is similar in absolute value to the contribution from $\hat{H}^{h/q}$.  
In fact those three contributions almost cancel each other leaving a nearly vanishing fragmentation piece. 
The resulting asymmetry is close to the contribution from  $F_{FT}^q$ and is larger than the experimental data, as clearly seen in the figure. 
The experimental data are around 5\% and our computations result in a positive asymmetry of about 15\%. 
 
\begin{figure}[h!t]
\includegraphics[width=8.truecm,angle=0]{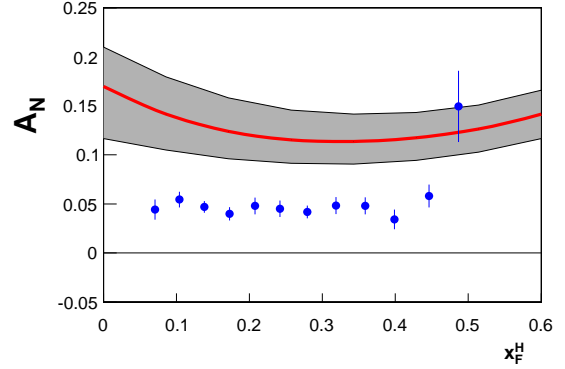}
\caption{$A_N$ as a function of $x_F^H$ for $\pi^+$  production at $\Php = 1 \; \rm{GeV}$ and $\sqrt{S} = 7.25 \; \rm {GeV}$. 
The data are from  Ref.~\cite{Airapetian:2013bim}.  
The solid line corresponds to sum of all contributions with $\hat{H}^{h/q,\Im}_{FU} = 0$.}
\label{fig:an_hermes_HF_pipm_xf}
\end{figure}

 Let us elaborate more on the contribution due to the 3-parton correlator $\hat{H}^{h/q,\Im}_{FU}$. 
According to Ref.~\cite{Kanazawa:2014dca}, $\hat{H}^{h/q,\Im}_{FU}$, in particular through its contribution to $H^{h/q}$ via Eq.~\eqref{e:relation}, might play a critical role for the description of $A_N$ in $p^{\uparrow} p \to h \, X$ in the collinear twist-3 approach.
In Fig.~\ref{fig:an_hermes_HF_pipm_xf} we present our computations for $A_N$ when $\hat{H}^{h/q,\Im}_{FU}$ is switched off.
(Note that setting $H^{h/q}$ and $\hat{H}^{h/q,\Im}_{FU}$ to zero simultaneously would also imply $\hat{H}^{h/q} = 0$ due to the relation in Eq.~\eqref{e:relation}.) 
We remind the reader that $\hat{H}^{h/q,\Im}_{FU}$ does not have an analogue among TMD functions, thus using  $\hat{H}^{h/q} = 0$ simplifies comparison with TMD formalism.
Comparing with Fig.~\ref{fig:an_hermes_pipm_xf} one observes that $A_N^{\pi^+}$ does not change very much. However, one must keep in mind that the function $\hat{H}^{h/q,\Im}_{FU}$ was fitted to experimental data in proton-proton scattering which are in the large positive $x_F$ range (i.e., large negative $x_F^{H}$ region) not explored by inclusive hadron production in lepton-proton scattering at HERMES. Error bands for these functions were not computed in Ref.~\cite{Kanazawa:2014dca}.

\subsection{Comparison and conclusions}

Here we give a brief  comparison between the collinear twist-3  approach described in Sec~\ref{sectionII} and the TMD Generalised Parton Model (GPM) described in Sec~\ref{sectionIII} from both a conceptual and a phenomenological point of view.
The GPM has been applied to $A_N$ for $\ell \, p^{\uparrow} \to h \, X$~\cite{Anselmino:1999gd,Anselmino:2009pn,Anselmino:2014eza} and for $p^{\uparrow} p \to h \, X$ --- see~\cite{Anselmino:1994tv,Anselmino:1998yz,Anselmino:2005sh,Anselmino:2012rq,Anselmino:2013rya} and references therein.
This model uses 2-parton correlation functions only, but consistently keeps the transverse parton momenta at all stages of the calculation. In the case of twist-3 observables like $A_N$, not all leading power terms are covered by the GPM.\footnote{A closely related discussion about the twist-3 so-called Cahn effect in SIDIS can be found in Ref.~\cite{Bacchetta:2008xw}.}
This holds for the twist-3 effect on the distribution side~\cite{Gamberg:2010tj} and, in particular, also for the twist-3 fragmentation contribution~\cite{Metz:2012ct}. 
As mentioned above, for the latter one has two independent fragmentation correlators~\cite{Metz:2012ct}, while in the GPM only the Collins function contributes.
(At present, a detailed analytical comparison of the fragmentation contributions in the two approaches does not exist.)
On the other hand, the GPM contains certain (kinematical) higher twist contributions and may also mimic effects of a collinear higher order calculation at leading twist. 
We note in passing that a recipe for incorporating in the GPM the process dependence of the Sivers effect~\cite{Collins:2002kn} has been discussed in~\cite{Gamberg:2010tj}.
 
Let us now turn to the phenomenology of $A_N$ for $\ell \, p^{\uparrow} \to h \, X$.
The GPM predictions are closer to the HERMES data than what we found in the collinear twist-3 framework, where the best results in the GPM were obtained by exploiting somewhat older extractions of the Sivers function and the Collins function~\cite{Anselmino:2005ea,Anselmino:2007fs} --- compare Fig.~1 and Fig.~2 in~\cite{Anselmino:2014eza} with our Fig.~\ref{fig:an_hermes_pipm_xf}.
However, one again has to keep in mind the aforementioned underestimated error of the twist-3 calculation and the need for a NLO calculation.
Moreover, due to large error bands, no conclusion could be drawn as to whether the Sivers or Collins effect can describe $A_N$ in $p^\uparrow p \to \pi X$ within the GPM~\cite{Anselmino:2012rq,Anselmino:2013rya}. 
In this regard, a much more definite statement was made with the collinear twist-3 analysis performed in Ref.~\cite{Kanazawa:2014dca}, i.e., that the fragmentation mechanism in that formalism can be the cause of the transverse single-spin asymmetries seen in pion production from proton-proton collisions.  

We find that twist-3 results with $\hat{H}^{h/q,\Im}_{FU} = 0$ in Fig.~\ref{fig:an_hermes_HF_pipm_xf} have the same signs and are close in magnitude to the curves labeled as SIDIS 2 in Figs.~1 and 2 of  Fig.~\ref{fig:an-hermes-pip}. 
One may speculate then that an analytical relation between the GPM and twist-3 approaches (showing where the two formalisms agree and/or differ) is perhaps possible for this observable if one neglects the 3-parton FF.  
However, as already stated, no such rigorous derivation has been performed yet.
Let us also mention that prediction for $A_N^{\pi^+}$ for the EIC of Ref.~\cite{Gamberg:2014eia} are comparable both in sign and size with those of Refs.~\cite{Anselmino:2008sga,Anselmino:2014eza} using GPM framework.  On the other hand, our result for $A_N^{\pi^-}$ for the EIC is quite different from what one finds in the GPM~\cite{Anselmino:2008sga,Anselmino:2014eza}. 
Such a measurement might therefore allow one to discriminate between the phenomenology of the two approaches. 

Further studies both theoretical and experimental are needed to clarify the issues of relation of two factorization mechanisms and the origin of the process. 
We emphasize the need for computing the NLO corrections and assess its impact on $A_N$, especially in the region of lower transverse hadron momenta $\Php$.
Moreover, we note the error of our numerical calculations in Ref.~\cite{Gamberg:2014eia} is underestimated in case of twist-3 calculations.
In this regard it will be important to better constrain the 3-parton fragmentation correlator $\hat{H}_{FU}^{h/q,\Im}$.
On the experimental side, it would be very useful to have absolute cross section measurements from both HERMES and Jefferson Lab, which would help one to obtain a quantitative understanding of the role played by higher order corrections.
%
%
\section*{Acknowledgments}
\noindent 
I would like to thank  Mauro Anselmino, Mariaelena Boglione,
Umberto D'Alesio, 
Stefano Melis,  
Francesco Murgia, Leonard Gamberg, 
 Zhong-Bo Kang, Andreas Metz, Daniel Pitonyak  for collaboration. The results of these proceedings belong to them and are based on  Refs.~\cite{Anselmino:2014eza,Gamberg:2014eia}.
 
This material is based upon work supported by the U.S. Department of Energy, 
Office of Science, Office of Nuclear Physics, under contract No.~No.~DE-AC05-06OR23177 (A.P.).

\bibliography{prokudin_transversity14}

\begin{thebibliography}{66}

\bibitem{Anselmino:2014eza}
M.~Anselmino, M.~Boglione, U.~D'Alesio, S.~Melis, F.~Murgia et~al., Phys.~Rev.
  \textbf{D89}, 114026 (2014), \texttt{1404.6465}

\bibitem{Gamberg:2014eia}
L.~Gamberg, Z.B. Kang, A.~Metz, D.~Pitonyak, A.~Prokudin, Phys.Rev.
  \textbf{D90}, 074012 (2014), \texttt{1407.5078}

\bibitem{Adams:1991cs}
D.L. Adams et~al. (FNAL-E704), Phys.~Lett. \textbf{B264}, 462 (1991)

\bibitem{Adler:2005in}
S.S. Adler et~al. (PHENIX), Phys.~Rev.~Lett. \textbf{95}, 202001 (2005),
  \texttt{hep-ex/0507073}

\bibitem{Adare:2013ekj}
A.~Adare et~al. (PHENIX Collaboration) (2013), \texttt{1312.1995}

\bibitem{Adare:2014qzo}
A.~Adare et~al. (PHENIX Collaboration) (2014), \texttt{1406.3541}

\bibitem{Adams:2003fx}
J.~Adams et~al. (STAR), Phys.~Rev.~Lett. \textbf{92}, 171801 (2004),
  \texttt{hep-ex/0310058}

\bibitem{Abelev:2008qb}
B.I. Abelev et~al. (STAR), Phys.~Rev.~Lett. \textbf{101}, 222001 (2008),
  \texttt{0801.2990}

\bibitem{Adamczyk:2012xd}
L.~Adamczyk et~al. (STAR Collaboration), Phys.~Rev. \textbf{D86}, 051101
  (2012), \texttt{1205.6826}

\bibitem{Arsene:2008mi}
I.~Arsene et~al. (BRAHMS), Phys.~Rev.~Lett. \textbf{101}, 042001 (2008),
  \texttt{0801.1078}

\bibitem{Bland:2013pkt}
L.~Bland et~al. (AnDY Collaboration) (2013), \texttt{1304.1454}

\bibitem{Efremov:1981sh}
A.V. Efremov, O.V. Teryaev, Sov.~J.~Nucl.~Phys. \textbf{36}, 140 (1982)

\bibitem{Efremov:1984ip}
A.V. Efremov, O.V. Teryaev, Phys.~Lett. \textbf{B150}, 383 (1985)

\bibitem{Qiu:1991pp}
J.W. Qiu, G.~Sterman, Phys.~Rev.~Lett. \textbf{67}, 2264 (1991)

\bibitem{Qiu:1991wg}
J.W. Qiu, G.~Sterman, Nucl.~Phys. \textbf{B378}, 52 (1992)

\bibitem{Qiu:1998ia}
J.W. Qiu, G.~Sterman, Phys.~Rev. \textbf{D59}, 014004 (1998),
  \texttt{hep-ph/9806356}

\bibitem{Kouvaris:2006zy}
C.~Kouvaris, J.W. Qiu, W.~Vogelsang, F.~Yuan, Phys.~Rev. \textbf{D74}, 114013
  (2006), \texttt{hep-ph/0609238}

\bibitem{Kang:2011hk}
Z.B. Kang, J.W. Qiu, W.~Vogelsang, F.~Yuan, Phys.~Rev. \textbf{D83}, 094001
  (2011), \texttt{1103.1591}

\bibitem{Anselmino:1994tv}
M.~Anselmino, M.~Boglione, F.~Murgia, Phys.Lett. \textbf{B362}, 164 (1995),
  \texttt{hep-ph/9503290}

\bibitem{Anselmino:1998yz}
M.~Anselmino, F.~Murgia, Phys.~Lett. \textbf{B442}, 470 (1998),
  \texttt{hep-ph/9808426}

\bibitem{Anselmino:2005sh}
M.~Anselmino, M.~Boglione, U.~D'Alesio, E.~Leader, S.~Melis et~al., Phys.~Rev.
  \textbf{D73}, 014020 (2006), \texttt{hep-ph/0509035}

\bibitem{Anselmino:2012rq}
M.~Anselmino, M.~Boglione, U.~D'Alesio, E.~Leader, S.~Melis et~al., Phys.~Rev.
  \textbf{D86}, 074032 (2012), \texttt{1207.6529}

\bibitem{Anselmino:2013rya}
M.~Anselmino, M.~Boglione, U.~D'Alesio, S.~Melis, F.~Murgia et~al., Phys.~Rev.
  \textbf{D88}, 054023 (2013), \texttt{1304.7691}

\bibitem{Kanazawa:2000hz}
Y.~Kanazawa, Y.~Koike, Phys.~Lett. \textbf{B478}, 121 (2000),
  \texttt{hep-ph/0001021}

\bibitem{Eguchi:2006mc}
H.~Eguchi, Y.~Koike, K.~Tanaka, Nucl.~Phys. \textbf{B763}, 198 (2007),
  \texttt{hep-ph/0610314}

\bibitem{Koike:2009ge}
Y.~Koike, T.~Tomita, Phys.~Lett. \textbf{B675}, 181 (2009), \texttt{0903.1923}

\bibitem{Kanazawa:2010au}
K.~Kanazawa, Y.~Koike, Phys.~Rev. \textbf{D82}, 034009 (2010),
  \texttt{1005.1468}

\bibitem{Kanazawa:2011bg}
K.~Kanazawa, Y.~Koike, Phys.~Rev. \textbf{D83}, 114024 (2011),
  \texttt{1104.0117}

\bibitem{Beppu:2013uda}
H.~Beppu, K.~Kanazawa, Y.~Koike, S.~Yoshida, Phys.~Rev. \textbf{D89}, 034029
  (2014), \texttt{1312.6862}

\bibitem{Kang:2010zzb}
Z.B. Kang, F.~Yuan, J.~Zhou, Phys.~Lett. \textbf{B691}, 243 (2010),
  \texttt{1002.0399}

\bibitem{Metz:2012ui}
A.~Metz, D.~Pitonyak, A.~Schaefer, M.~Schlegel, W.~Vogelsang et~al., Phys.~Rev.
  \textbf{D86}, 094039 (2012), \texttt{1209.3138}

\bibitem{Metz:2012ct}
A.~Metz, D.~Pitonyak, Phys.~Lett. \textbf{B723}, 365 (2013), \texttt{1212.5037}

\bibitem{Kanazawa:2014dca}
K.~Kanazawa, Y.~Koike, A.~Metz, D.~Pitonyak (2014), \texttt{1404.1033}

\bibitem{Airapetian:2013bim}
A.~Airapetian et~al. (HERMES Collaboration), Phys.~Lett. \textbf{B728}, 183
  (2014), \texttt{1310.5070}

\bibitem{Allada:2013nsw}
K.~Allada et~al. (Jefferson Lab Hall A Collaboration), Phys.~Rev. \textbf{C89},
  042201 (2014), \texttt{1311.1866}

\bibitem{Sivers:1989cc}
D.W. Sivers, Phys.~Rev. \textbf{D41}, 83 (1990)

\bibitem{Sivers:1990fh}
D.W. Sivers, Phys.~Rev. \textbf{D43}, 261 (1991)

\bibitem{Yuan:2009dw}
F.~Yuan, J.~Zhou, Phys.~Rev.~Lett. \textbf{103}, 052001 (2009),
  \texttt{0903.4680}

\bibitem{Kanazawa:2013uia}
K.~Kanazawa, Y.~Koike, Phys.~Rev. \textbf{D88}, 074022 (2013),
  \texttt{1309.1215}

\bibitem{Collins:1992kk}
J.C. Collins, Nucl.~Phys. \textbf{B396}, 161 (1993), \texttt{hep-ph/9208213}

\bibitem{Koike:2002ti}
Y.~Koike, AIP Conf.Proc. \textbf{675}, 449 (2003), \texttt{hep-ph/0210396}

\bibitem{Anselmino:1999gd}
M.~Anselmino, M.~Boglione, J.~Hansson, F.~Murgia, Eur.~Phys.~J. \textbf{C13},
  519 (2000), \texttt{hep-ph/9906418}

\bibitem{Anselmino:2009pn}
M.~Anselmino, M.~Boglione, U.~D'Alesio, S.~Melis, F.~Murgia et~al., Phys.~Rev.
  \textbf{D81}, 034007 (2010), \texttt{0911.1744}

\bibitem{D'Alesio:2004up}
U.~D'Alesio, F.~Murgia, Phys. Rev. \textbf{D70}, 074009 (2004),
  \texttt{hep-ph/0408092}

\bibitem{D'Alesio:2007jt}
U.~D'Alesio, F.~Murgia, Prog. Part. Nucl. Phys. \textbf{61}, 394 (2008),
  \texttt{0712.4328}

\bibitem{Kang:2011jw}
Z.B. Kang, A.~Metz, J.W. Qiu, J.~Zhou, Phys.~Rev. \textbf{D84}, 034046 (2011),
  \texttt{1106.3514}

\bibitem{Kang:2008qh}
Z.B. Kang, J.W. Qiu, Phys.~Rev. \textbf{D78}, 034005 (2008), \texttt{0806.1970}

\bibitem{Kang:2008ih}
Z.B. Kang, J.W. Qiu, W.~Vogelsang, F.~Yuan, Phys.~Rev. \textbf{D78}, 114013
  (2008), \texttt{0810.3333}

\bibitem{Zhou:2009jm}
J.~Zhou, F.~Yuan, Z.T. Liang, Phys.~Rev. \textbf{D81}, 054008 (2010),
  \texttt{0909.2238}

\bibitem{Beppu:2010qn}
H.~Beppu, Y.~Koike, K.~Tanaka, S.~Yoshida, Phys.~Rev. \textbf{D82}, 054005
  (2010), \texttt{1007.2034}

\bibitem{Koike:2011mb}
Y.~Koike, S.~Yoshida, Phys.~Rev. \textbf{D84}, 014026 (2011),
  \texttt{1104.3943}

\bibitem{Liang:2012rb}
Z.T. Liang, A.~Metz, D.~Pitonyak, A.~Schaefer, Y.K. Song et~al., Phys.~Lett.
  \textbf{B712}, 235 (2012), \texttt{1203.3956}

\bibitem{Metz:2012fq}
A.~Metz, D.~Pitonyak, A.~Schaefer, J.~Zhou, Phys.~Rev. \textbf{D86}, 114020
  (2012), \texttt{1210.6555}

\bibitem{Hatta:2013wsa}
Y.~Hatta, K.~Kanazawa, S.~Yoshida, Phys.~Rev. \textbf{D88}, 014037 (2013),
  \texttt{1305.7001}

\bibitem{Boer:2003cm}
D.~Boer, P.J. Mulders, F.~Pijlman, Nucl.~Phys. \textbf{B667}, 201 (2003),
  \texttt{hep-ph/0303034}

\bibitem{Bacchetta:2004jz}
A.~Bacchetta, U.~D'Alesio, M.~Diehl, C.A. Miller, Phys.~Rev. \textbf{D70},
  117504 (2004), \texttt{hep-ph/0410050}

\bibitem{Anselmino:2005ea}
M.~Anselmino, M.~Boglione, U.~D'Alesio, A.~Kotzinian, F.~Murgia et~al.,
  Phys.~Rev. \textbf{D72}, 094007 (2005), \texttt{hep-ph/0507181}

\bibitem{Kretzer:2000yf}
S.~Kretzer, Phys. Rev. \textbf{D62}, 054001 (2000), \texttt{hep-ph/0003177}

\bibitem{Anselmino:2008sga}
M.~Anselmino, M.~Boglione, U.~D'Alesio, A.~Kotzinian, S.~Melis et~al.,
  Eur.~Phys.~J. \textbf{A39}, 89 (2009), \texttt{0805.2677}

\bibitem{deFlorian:2007aj}
D.~de~Florian, R.~Sassot, M.~Stratmann, Phys.~Rev. \textbf{D75}, 114010 (2007),
  \texttt{hep-ph/0703242}

\bibitem{Anselmino:2007fs}
M.~Anselmino, M.~Boglione, U.~D'Alesio, A.~Kotzinian, F.~Murgia et~al.,
  Phys.Rev. \textbf{D75}, 054032 (2007), \texttt{hep-ph/0701006}

\bibitem{Anselmino:2008jk}
M.~Anselmino, M.~Boglione, U.~D'Alesio, A.~Kotzinian, F.~Murgia et~al.,
  Nucl.Phys.Proc.Suppl. \textbf{191}, 98 (2009), \texttt{0812.4366}

\bibitem{Anselmino:2013vqa}
M.~Anselmino, M.~Boglione, U.~D'Alesio, S.~Melis, F.~Murgia et~al., Phys.~Rev.
  \textbf{D87}, 094019 (2013), \texttt{1303.3822}

\bibitem{Bacchetta:2008xw}
A.~Bacchetta, D.~Boer, M.~Diehl, P.J. Mulders, JHEP \textbf{08}, 023 (2008),
  \texttt{0803.0227}

\bibitem{Gamberg:2010tj}
L.~Gamberg, Z.B. Kang, Phys.~Lett. \textbf{B696}, 109 (2011),
  \texttt{1009.1936}

\bibitem{Collins:2002kn}
J.C. Collins, Phys.~Lett. \textbf{B536}, 43 (2002), \texttt{hep-ph/0204004}

\end{thebibliography}

\end{document}